# Photomolecular Effect as A Potential Explanation for The Cloud Absorption Anomaly

Gang Chen[*]

Department of Mechanical Engineering

Massachusetts Institute of Technology

Cambridge, MA 02139

**Abstract**

Cloud absorption is acknowledged as the biggest source of uncertainty in the climate models. For over 70 years, many experiments have reported clouds absorbing more solar radiation than theory could predict. In the visible spectrum, simulations based on optical constants of water lead to negligible cloud absorption. This result had been explored by some experimentalists to calibrate the cloud absorption measurements. However, the author and his collaborators recently discovered that visible light can directly cleave off water molecular clusters at liquid-air interfaces (PNAS, e2312751120, 2023; e2320844121, 2024), which is named the photomolecular effect in analogy to the photoelectric effect. This discovery suggests that a crucial piece of physics has been missing in the existing theories: light can be absorbed at water-air interface. The photomolecular effect can be simulated by generalizing the boundary conditions for the Maxwell equations using Feibelman parameters that were derived in the past research on the surface photoelectric effect and surface plasmons. In this work, the author uses simulation to show that including the photomolecular effect at the air-water interface can potentially explain the cloud absorption anomaly. Although the lack of accurate Feibelman parameter values prevents a direct comparison of the current theory with experiments, this work points to an unexplored mechanism for explaining the cloud absorption anomaly and calls for further investigation on the impacts of photomolecular effect in the climate modeling.

---

[*] Email: gchen2@mit.edu

**Significance Statement**:

For over 70 years, some experiments have reported clouds absorb more sunlight than what theory could predict. Such anomalous clouds absorption has significant implications on the earth's water cycle and climate modeling, and hence its existence has been subject to significant scrutiny and heated debate. This work suggests that the recently discovered photomolecular effect - photons in the visible frequency range directly cleave off water molecular clusters at the air-water interface – provides a new mechanism to explain the anomalous cloud absorption. The work establishes a simulation approach to assess the impacts of this effect, and the simulation results confirm the potential of the photomolecular effect to explain the clouds absorption anomaly.

1. **Introduction**

Discrepancies between experiments and models of clouds absorption have long existed. Such uncertainties are widely acknowledged as the most uncertain part of climate models (Stephens and Tsay 1990; Valero et al. 2004; Wild et al. 1995; Trenberth et al. 2009). In 1951, Fritz and McDonald(1951) used flight measurements above the clouds and pyrheliometer data on the ground to deduce that the measured absorptance values of clouds ranging 13-27% in three different areas were significantly higher than that of the calculation by Hewson(Hewson 1943) which does not exceed 6%. In subsequent 74 years, many experiments have reported clouds absorbing more solar radiation than theory could calculate (Cess et al. 1995; Ramanathan et al. 1995; Pilewskie and Valero 1995; O'Hirok and Gautier 2003; Schmidt et al. 2010; Lee et al. 2021; Wild et al. 1995; Zhang et al. 1997). However, due to the complexity of the cloud structures and aerosols, the existence of such cloud absorption anomaly has been disputed (Stephens 1996; Stephens and Tsay 1990; Li et al. 2003, 2004, 1995; Valero et al. 2004). Stephen and Tsai (1990) postulated two explanations for this anomaly (1) aerosol absorption and droplet size uncertainties, and (2) spatial inhomogeneity invalidating the one-dimensional model typically used to explain the experimental data. Due to difficulties in quantify aerosols and cloud inhomogeneities, the discrepancies between experiments and models remain unsolved.

In the visible spectrum, bulk water absorbs little solar radiation. The penetration depth of green light in liquid water is ~40 m. A 1 km thick cloud with liquid water path of 1000 g/m$^2$ contains equivalent of 1 m thick liquid water. Despite multiple scattering of water droplets in clouds that effectively increases the light path length, past simulations based on the optical constants of water have shown that cloud absorption in the visible spectrum is negligible (Stephens and Tsay 1990). This fact has been used as an invariance to calibrate experiments or conditioning experimental data. For example, Valero et al. (2003) assumed that the 500 nm absorption is invariant with the cloud amount. Ackerman and Cox used a single wavelength irradiance measurement at 500 nm to correct broadband measurement of the cloud flux divergence. Oreopoulos et al.(2003) used such invariance for conditioned sampling of their experimental data above and below the clouds. The same strategy was used by Kindel et al. (2011) in analyzing spectral cloud absorptance data.

However, some experiments have suggested clouds do absorb in the visible spectrum. For example, in the atmospheric-radiation-measurement enhanced-shortwave-experiment program, both broad band (0.0224-3.91 μm) and near infrared band (0.68-3.3 μm) solar flux data were measured with synchronized flights above and below clouds. Analysis of the albedo and absorptance data suggests that absorption happen in the visible spectrum (Zender et al. 1997). Follow up experiments five years later also reported anomalous absorption at 500 nm (Oreopoulos et al. 2003). Spectroscopic measurement of atmospheric absorption clearly show absorption in the visible region in some cases, even after considering the cloud inhomogeneity via modeling (Schmidt et al. 2010).

Recent experiments from the author and his collaborators show that visible light is actually absorbed at water-air interface in a process named as the photomolecular effect: visible photons directly cleave off water molecular clusters (Tu et al. 2023; Lv et al. 2024). The photomolecular effect was conceived to explain super-thermal solar interfacial-evaporation rates observed in some porous materials such as hydrogels and was further confirmed by experiments at a single air-water interface (Tu et al. 2023; Lv et al. 2024; Zhao et al. 2018; Zhang et al. 2025). The driving force for the interfacial absorption process was believed to be due to the large electrical field gradient at the interface as dictated by the continuity of the displacement field in the direction perpendicular to the interface, which leads to a net force acting on the water molecules at interfaces (Tu et al. 2023; Lv et al. 2024; Landry et al. 2025). At a single air-water interface, some of the observed phenomena supporting the photomolecular effect are: (1) green laser heats up water and the temperature rise depends on polarization and angle of incidence: several times higher for the transverse-magnetic (TM) than for transverse-electric (TE) polarization and peaking at 45º of incidence, (2) both pump-and-probe mirage experiments on the vapor side and laser refraction experiments through air-water interfaces show strong response to TM-polarization and at 45º of incidence, peaking at green light, (3) air above the water surface cools down under TM laser irradiation followed by a saturation region, due to water clusters breaking up in air, and (4) Raman and IR spectra show signs of cluster existence in the air. Furthermore, it was shown that LEDs with a power at 1000 W/m$^2$ heat up a fog chamber. Surprisingly, the photomolecular effect is found to be strongest at green light, where water is least absorbing. Different experiments carried out by Verma et al. (2024) recently validated some of these original observations. The photomolecular effect also provides a potential explanation to previously observed breathing response of ultrathin water layer adsorbed on diamond to visible light (Sommer et al. 2011). The latter observation was actually suspected by some to be the reason behind many observed therapeutic effects of low-level light on different diseases (Sommer et al. 2020).

The photomolecular effect provides a potentially new mechanism to explain the anomalous cloud absorption: interfacial absorption. Although this discovery has yet to be widely confirmed and a microscopic theory is just emerging (Landry et al. 2025), it is an interfacial process at the air-liquid water interface that is abundant in clouds. The author had recently shown (Chen 2024) that the generalized boundary conditions of the Maxwell equations using the Feibelman parameters to describe the surface photoelectric effect and surface plasmons (Feibelman 1982; Liebsch 1997; Zhen et al. 2013; Gonçalves et al. 2020; Schaich and Chen 1989) are also applicable for treating the photomolecular effect, which means that corresponding modifications of the Mie scattering theory for metallic nanoparticles (Gonçalves et al. 2020) should also be valid for water droplets. Hence, the mathematical basis to include the photomolecular effect into cloud modeling already exists. Although accurate values of the Feibelman parameters at the air-water interface and their wavelength dependence are not known yet, it will be useful to assess the potential impacts of this new effect.

This paper presents an assessment of potential impacts of the photomolecular effect for cloud absorption. The author will start from the modified solution of the Mie scattering theory using the Feibelman parameters to calculate the scattering and absorption efficiencies of a single water droplet, followed by proper averaging of these properties based on the cloud particle distribution to obtain the absorption and extinction coefficients and phase function as the input for the equation of radiative transfer (ERT). An inhouse code was developed following the established procedure solving the ERT (Liou 1973; Stephens 1978; Stamnes et al. 1988) and applied to obtain numerical results. Despite the uncertainties in the Feibelman parameters, the simulation results show that the photomolecular effect can potentially explain the anomalies of cloud absorption.

## 2. Formulation and Results

a. **Generalized Boundary Conditions for Maxwell Equations**.

The established boundary conditions for the Maxwell equations require that the perpendicular displacement field across the interface is continuous, i.e., $D_z(0^+) = D_z(0^-)$, where z denotes the direction perpendicular to the interface. This implies that a discontinuity of the electrical field exists across the interface. In reality, however, the electrical field is just a steeply changing function rather than discontinuous. Using this picture, one can derive a new set of boundary conditions for the Maxwell equations as follows (Feibelman 1982; Liebsch 1997; Zhen et al. 2013; Chen 2024; Schaich and Chen 1989; Lai et al. 2021; Gonçalves et al. 2020)

$$[\![\mathbf{H}_\|]\!] = i\omega d_\| [\![\mathbf{D}_\|]\!] \times \mathbf{n}_{12} \qquad (1)$$

$$[\![\mathbf{D}_\perp]\!] = d_\| \nabla_\| \cdot [\![\mathbf{D}_\|]\!] \qquad (2)$$

$$[\![\mathbf{E}_\|]\!] = -d_\perp \nabla_\| [\![E_\perp]\!] \qquad (3)$$

$$[\![\mathbf{B}_\perp]\!] = 0 \qquad (4)$$

where the operator $[\![\ ]\!]$ means the difference across the interface for the quantity inside the bracket, i.e., $[\![D_z]\!] = D_z(0^+) - D_z(0^-)$. The subscript "$\|$" means parallel to the interface, while "$\perp$" means perpendicular to the interface, $\mathbf{n}_{12}$ is the unit vector pointing from medium 1 (air side) to medium 2 (water side). The gradient operator $\nabla_\|$ is taken parallel to the local surface tangent and is equal to $\left(\hat{x}\frac{\partial}{\partial x} + \hat{y}\frac{\partial}{\partial y}\right) = \hat{x} i k_x$ at a flat interface. In the above expressions, $d_\|$ and $d_\perp$ are the Feibelman parameters defined as

$$d_\perp = \frac{\int_{z_1}^{z_2} z\frac{dE_z}{dz}dz}{\int_{z_1}^{z_2} \frac{dE_z}{dz}dz} = \frac{\int_{z_1}^{z_2} z\rho(z)dz}{\int_{z_1}^{z_2} \rho(z)dz} = \mathbf{d}_\perp \cdot \mathbf{n}_{12} \qquad (5)$$

$$d_\| = \frac{\int_{z_1}^{z_2} z\frac{dD_x}{dz}dz}{\int_{z_1}^{z_2} \frac{dD_x}{dz}dz} = \frac{\int_{z_1}^{z_2} z\frac{\partial P_x}{\partial z}dz}{\int_{z_1}^{z_2} \frac{\partial P_x}{\partial z}dz} = \frac{\int_{z_1}^{z_2} z\frac{\partial J_x}{\partial z}dz}{\int_{z_1}^{z_2} \frac{\partial J_x}{\partial z}dz} = \mathbf{d}_\| \cdot \mathbf{n}_{12} \qquad (6)$$

The real part of $d_\perp$ measures the spatial deviation of the induced surface charge from the interface, and the real part of $d_\parallel$ is a measure of the location of the gradient of induced polarization parallel to the interface, i.e., the Feibelman parameters are the surface response functions. The imaginary parts of the Feibelman parameters are related to dissipation. Vectors in Eqs.(5) and (6) emphasize that the Felbelman parameters have directionality, as they denote the locations of the surface charge and current from the interface. This point is important in the modification of the Mie theory as EM waves across the interface in both directions, i.e., in and out of the droplet.

b. **Modification to the Mie Theory Including the Photomolecular Effect.**

Goncalves et al.(2020) had revised the Mie theory using generalized boundary conditions Eqs.(1)-(4) for plasmonic particles. In the chosen coordinates and symbols, their solutions give the scattering ($Q_{sca}$) and extinction ($Q_{ext}$) efficiencies as

$$Q_{sca} = \frac{2}{x^2}\sum_{n=1}^{\infty}(2n+1)(|a_n|^2 + |b_n|^2) \tag{7}$$

$$Q_{ext} = \frac{2}{x^2}\sum_{n=1}^{\infty}(2n+1)Re(a_n + b_n) \tag{8}$$

$$a_n = \frac{m^2 j_n(mx)\psi_n'(x) - \psi_n'(mx)j_n(x) - (m^2-1)[j_n(mx)j_n(x)\bar{d}_{\perp n} + \psi_n'(mx)\psi_n'(x)\,\bar{d}_\parallel]}{m^2 j_n(mx)\xi_n'(x) - \psi_n'(mx)h_n^{(1)}(x) - (m^2-1)[h_n^{(1)}(x)j_n(mx)\bar{d}_{\perp n} + \xi_n'(x)\psi_n'(mx)\,\bar{d}_\parallel]} \tag{9}$$

$$b_n = \frac{j_n(mx)\psi_n'(x) - \psi_n'(y)j_n(x) - (m^2-1)x^2 j_n(mx)j_n(x)\,\bar{d}_\parallel}{j_n(mx)\xi_n'(x) - \psi_n'(mx)h_n^{(1)}(x) - (m^2-1)x^2 h_n^{(1)}(x)j_n(mx)\,\bar{d}_\parallel} \tag{10}$$

where $j_n(x)$ and $h_n^{(1)}$ are spherical Bessel and Hankel functions, and $\psi_n(x) \equiv x j_n(x)$ and $\xi_n(x) \equiv x h_n^{(1)}(x)$ are Riccati-Bessel functions and the prime means derivative, $m^2 = \frac{\varepsilon_2}{\varepsilon_1}$ and $x = \frac{2\pi R}{\lambda_o}$ is the size parameter with $\lambda_o$ the wavelength in air, and

$$\bar{d}_{\perp n} = n(n+1)\frac{d_\perp}{R} \text{ and } \bar{d}_\parallel = \frac{d_\parallel}{R} \tag{11}$$

Note that in Eqs. (9) and (10), the signs in front of the terms involving d-parameters are opposite to that of Goncalves et al.'s solutions since here surface norm is chosen from external medium into the droplet, to be consistent with the previous choice for a flat interface. The scattering phase function is (Craig F. Bohren and Donald R. Huffman 1998)

$$\Phi(\Theta) = 2\frac{|S_1|^2 + |S_2|^2}{x^2 Q_{sca}} \tag{12}$$

where

$$S_1 = \sum_{n=1}^{\infty} \frac{2n+1}{n(n+1)} [a_n \pi_n(cos\Theta) + b_n \tau_n(cos\Theta)] \tag{13}$$

$$S_2 = \sum_{n=1}^{\infty} \frac{2n+1}{n(n+1)} [b_n \pi_n(cos\Theta) + a_n \tau_n(cos\Theta)] \tag{14}$$

$$\pi_n(cos\Theta) = \frac{dP_n(cos\Theta)}{dcos\Theta} = \frac{2n-1}{n-1} cos\Theta \, \pi_{n-1}(cos\Theta) - \frac{n}{n-1} \pi_{n-2}(cos\Theta) \tag{15}$$

$$\tau_n(cos\Theta) = n \, cos\Theta \, \pi_n(cos\Theta) - (n+1)\pi_{n-1}(cos\Theta)$$

$$\pi_0 = 0; \, \pi_1 = 1; \, \pi_2 = 3cos\Theta \tag{17}$$

$$\tau_0 = 0; \, \tau_1 = cos\Theta; \, \tau_2 = 3cos(2\Theta) \tag{18}$$

where Θ represents the angle formed between the incident and the scattered angle. A Mie scattering code was written using the above expressions, by modifying Baldi's Mie codes(Baldi 2024).

c. **Averaged Cloud Properties.**

A gamma distribution is used for the cloud droplet radius distribution (Stephens and Tsay 1990) to obtain the average phase function, the absorption and scattering coefficients,

$$n(r) = \frac{N_o}{\Gamma(j)r_m} \left(\frac{r}{r_m}\right)^{j-1} exp\left(-\frac{r}{r_m}\right) \tag{19}$$

where $N_o$ is the total (volume) concentration of droplets and $\Gamma$ the gamma function, $r_m$ is a characteristic radius of the distribution. With this distribution, the effective radius is $r_e = (j+2)r_m$. Following Stephen and Tsay (1990), the following parameters are taken: j=7 and $r_m$ =0.677 μm so that $r_e$=6.093 μm. This distribution function leads to cloud's water content w (kg/m³).

$$w = \frac{4}{3}\pi \rho_w N_o r_m^3 f(3) \tag{20}$$

where $f(m) = \Gamma(j+m)/\Gamma(j)$, and the liquid water path is the $L_p = wL$, where L is the cloud thickness. The volumetric extinction and absorption coefficients are calculated from

$$\alpha_e = \pi \int_0^\infty n(r) Q_e r^2 dr \quad \text{and} \quad \alpha_a = \pi \int_0^\infty n(r) Q_a r^2 dr \tag{21}$$

d. **Solution of the Equation of Radiative Transfer.**

The absorption and extinction coefficients and phase function are the inputs for the equation of radiative transfer (ERT), for which different solution methods developed (Liou 1973; Stephens 1978; Stamnes et al. 1988; Chandrasekhar 1960; Modest 2013). Since the focus here is on assessing the potential impacts of the photomolecular effect on the cloud absorption of solar radiation, the mid infrared thermal radiation transfer is neglected. Furthermore, only the case of the normal incidence of solar radiation to the cloud layer is included. Also neglected is the aerosol and the vapor phase absorption, since the main purpose is to show the potential impacts of photomolecular absorption on the cloud properties, rather than seeking agreements between simulations and experiments due to uncertainties in the Feibelman parameter values. With these simplifications, the radiation intensity is split into a direct beam and a diffuse beam, as in standard procedures and the diffuse beam obeys the ERT (Modest 2013)

$$\mu \frac{dI(\zeta,\mu)}{d\zeta} = -I + \frac{\omega}{2}\int_{-1}^{1} I(\zeta,\mu')\Phi(\mu' \to \mu)d\mu' + \frac{\omega I_o}{2} \Phi(0 \to \mu) exp(-\zeta) \tag{22}$$

where µ=cosθ is the directional cosine, $\zeta = \alpha_e z$ is the nondimensional coordinate, and $\omega = \alpha_s/\alpha_e$ is the scattering albedo. Note here the coordinate system is chosen to align the positive z-direction along the solar beam direction such that θ=0, i.e., µ=1, represents the incoming solar beam, which explains the slight differences of Eq.(22) from that of typically seen in literature that marked the z-direction opposite to the incident beam direction (Chandrasekhar 1960; Stamnes 1986; Liou 1973). The direct beam is a solution for the ERT including out-scattering only,

$$I_c(\zeta, 0) = I_o \exp(-\zeta) \qquad (23)$$

with $I_o$ denoting the incoming solar radiation intensity. The total intensity is the summation of the direct and the diffuse intensities,

$$I_t(\zeta, \mu) = I_c(\zeta, 0)\delta(\mu) + I(\zeta, \mu) \qquad (24)$$

All the intensities above are spectral based. Matlab codes were developed for Eq.(22) following procedures as described in the popular code DISORT (Stamnes et al. 2000, 1988; Stamnes 1986).

## 3. Results and Discussion
### a. Single Droplet Properties.

Figure1(a) shows the calculated absorption efficiency as a function of the diameter when the bulk water does not absorb, for specific Feibelman parameters given in the figure. In the small diameter range, the absorptance shows strong oscillations due to interference effect. Figure 1(b) shows the extinction efficiency, which is much larger than the absorption efficiency since scattering is dominate. Figure1c illustrates the ratio of absorption and extinction efficiency, i.e., co-albedo. In the small particle regime, the surface absorption is more importance due to reduced scattering. Figure1(d) displays the surface and bulk absorption, plotted using the size parameter x as a variable. However, it should be reminded that when surface absorption is included, it is not just the size parameter, but also the ratio $\frac{d_\perp}{R}$ and $\frac{d_\parallel}{R}$ that contributes to the absorption. Hence, the wavelength used in the calculations is also given as in Fig.1(b). These figures show that bulk absorption increases with the diameter (size parameter) while the surface absorption decreases slightly (due to decreasing $\frac{d_\perp}{R}$ and $\frac{d_\parallel}{R}$). Also shown in Fig.1(d) is the case when both the surface and the bulk absorption exist, the total absorptance is close to a superposition of the two. Although in theory, the real part of the Feibelman parameters could impact the absorption, numerical testing between $d_\perp = (0, -1.84 \times 10^{-9})$ and $d_\perp = (1.0 \times 10^{-9}, -1.84 \times 10^{-9})$ shows negligible difference in absorptance.

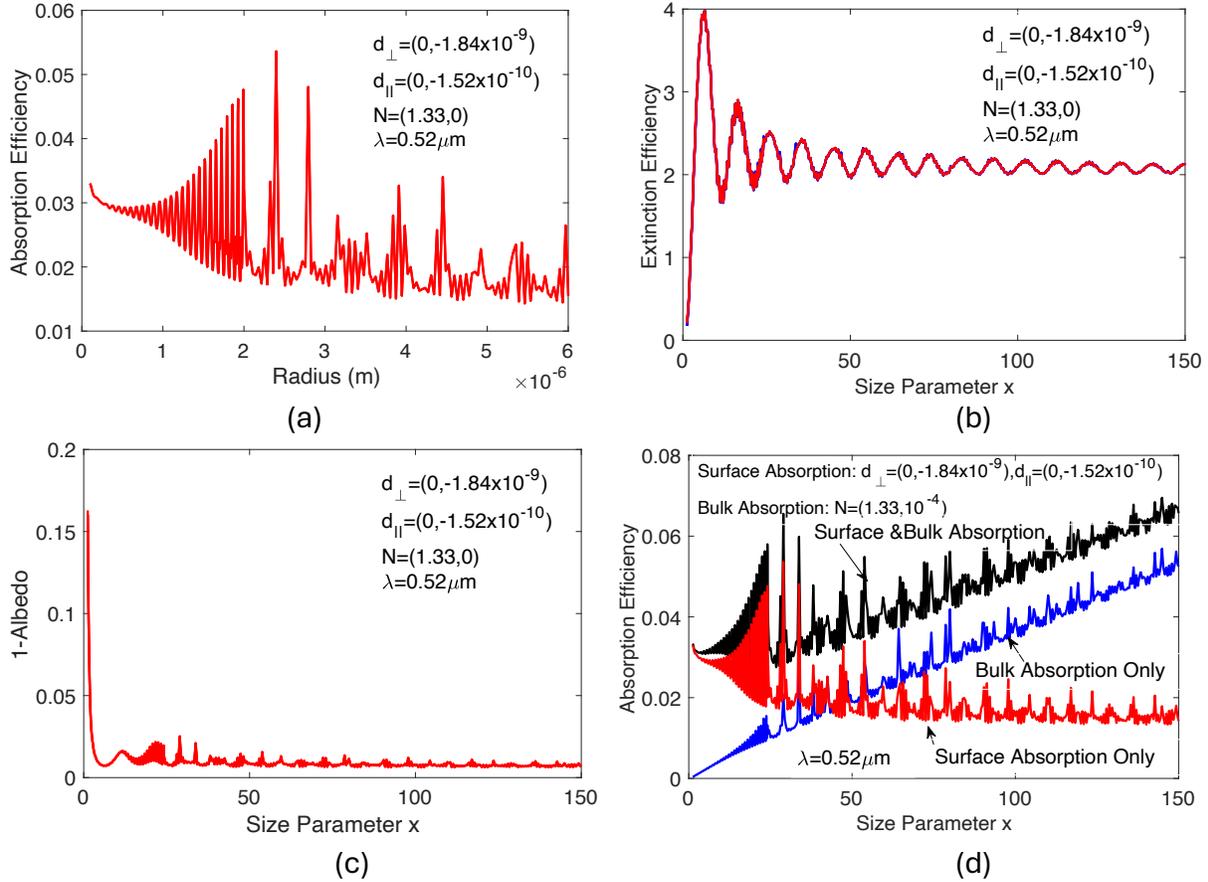

**Figure 1 Single droplet properties.** (a) Absorption efficiency as a function of radius for surface absorption only, (b) extinction efficiency and (c) co-albedo plotted as a function of the size parameter, which can be converted into diameter using the given wavelength, (d) comparion of absorption efficiency for surface or bulk absorption only, and when both surface and bulk absorption exist.

**b. Cloud Average Properties.**

Figures 2(a) shows the averaged absorption coefficient as a function of the effective droplet radius $r_e$, for fixed liquid water content. The surface absorption changes more rapidly with the effective radius because for fixed liquid water path, the smaller diameter means higher droplet density and more surface area. The albedo is shown in Fig.2(b). When no photomolecular effect exists, the albedo decreases with increasing radius due to increased bulk absorption. With the photomolecular effect, the albedo generally increases with effective radius, although a minimum was observed around 1 micron. Figures 2(c) and 2(d) illustrate the sensitivity of albedo to the Feibelman parameters. Although the albedo changes more rapidly with the imaginary part of $d_\parallel$, the current understanding of the photomolecular effect, as well as experiments, suggests that $d_\perp$ is much larger than $d_\parallel$ (Chen 2024; Landry et al. 2025; Liebsch 1997; Feibelman 1982). Figure 2(e) displays the phase function for a specific effective particle radius. As is typically the case for water droplets, most scattering is in the forward direction. Details of the phase function in other directions are shown in Fig.2(f).

For a planar surface, the surface absorptance depends only on the imaginary part.(Chen 2024) By fitting the estimated absorptance, the author inferred that the imaginary part of $d_\perp$ is ~ -10$^{-9}$ m and d$_{//}$ is ~ -10$^{-10}$ m. Danielson et al.(1969) had estimated before that the requirement co-albedo of clouds should be of the order 10$^{-3}$ to explain the anomalous cloud absorption, which is 10$^4$ order of magnitude larger than the calculated water droplet co-albedo in the visible spectrum based on bulk optical constants. Figure 2(c) shows that a co-albedo 10$^{-3}$ can be easily reached in the given Feibelman parameter range. In fact, values interfered before might be too high based on the calculated clouds transmittance results to be shown later. Furthermore, Landry et al.'s model (2025) suggests that the absorptance depends only weakly on wavelength in the visible spectrum, despite that the evaporation rate peaks at the green light. Due to these uncertainties, the Feibelman parameters at the air-water interface will be treated as adjustable and dependent of wavelength at this stage.

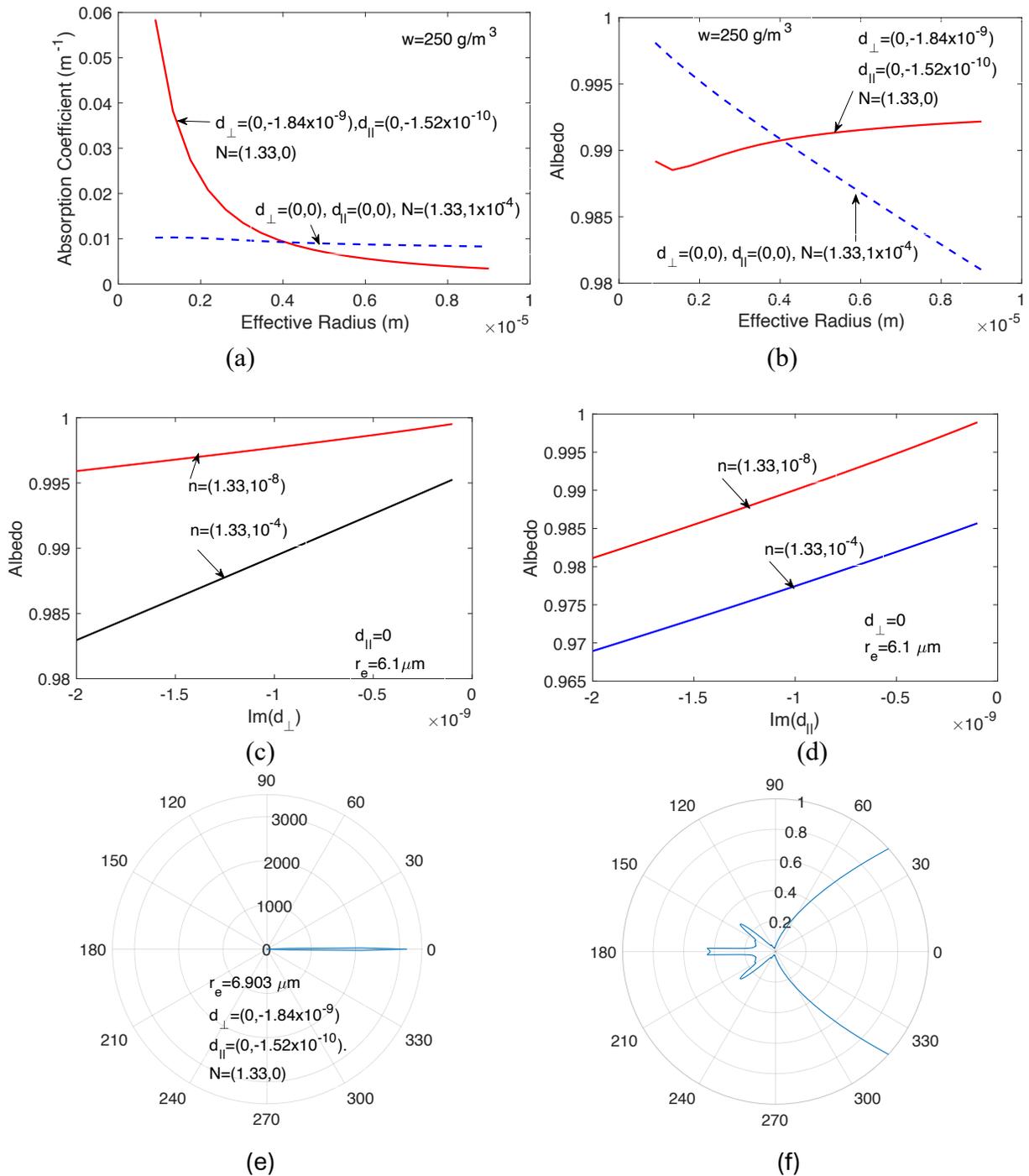

**Figure 2 Average cloud properties.** (a) Absorption efficiency and (b) albedo as a function of the effective droplet radius, (c) and (d) albedo sensitivity to the Feibelman parameters, (e) phase function for the given parameters and (f) enlarge portion of (e).

### c. Impacts of Surface Absorption on Cloud Properties.

Figures 3(a)-(c) show the cloud reflectance, i.e., albedo, diffuse transmittance, and absorptance as a function of wavelength for specified liquid water content $L_p$ and the average droplet diameter with different values of the $d_\perp$ parameter while setting $d_\parallel$ equaling zero. $d_\parallel$ is set to zero since experiments and theory suggests that $d_\perp$ is more important than $d_\parallel$ (Feibelman 1982; Liebsch 1997; Lv et al. 2024; Chen 2024; Landry et al. 2025). The water optical constants are taken from (Hale and Querry 1973). The direct transmittance is nearly zero in all wavelength range for the given parameters and hence is not shown. When both $d_\perp = d_\parallel = 0$, no surface absorption is included. In this case, solar radiation is mainly absorbed in the infrared region [Fig.3(c)]. However, when $d_\perp$ is not zero, finite absorption happens across the visible spectrum and in the near infrared region. As expected, larger values of $d_\perp$ leads to larger absorption in the visible and near infrared region. For wavelength over 1.3 μm, the surface absorption effect is small compared to bulk absorption.

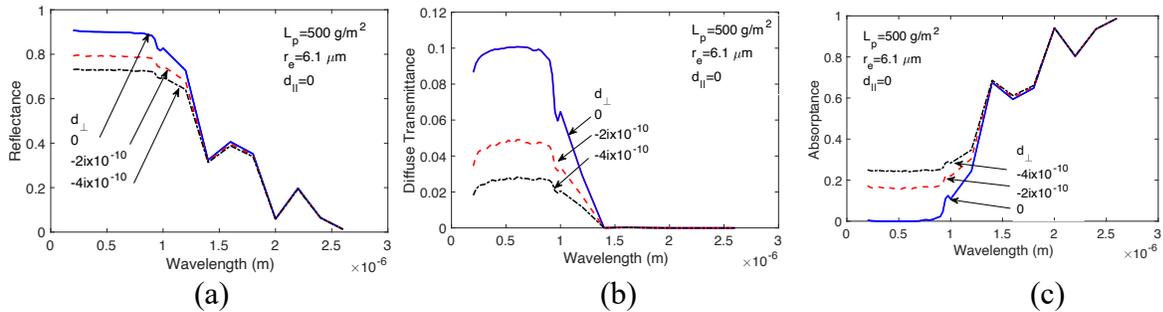

(a)                  (b)                  (c)

**Figure 3 Cloud properties: wavelength dependence.** (a) reflectance (albedo), (b) diffuse transmittance, and (c) absorptance as a function of wavelength for the given parameters.

Figures 4(a)-(c) show the cloud absorptance and other properties as a function of the liquid water path $L_p$ at $\lambda=0.52$ μm where bulk water does not absorb, for the given effective radius and different values of $d_\perp$ parameter. As expected, with increasing liquid water path, the absorptance increases due to the photomolecular effect, and concurrently, the albedo, i.e., reflectance, decreases. Such trends are expected if there exists absorption in clouds, and was used by Zender et al.(Zender et al. 1997) to support their claim of absorption in the visible spectrum.

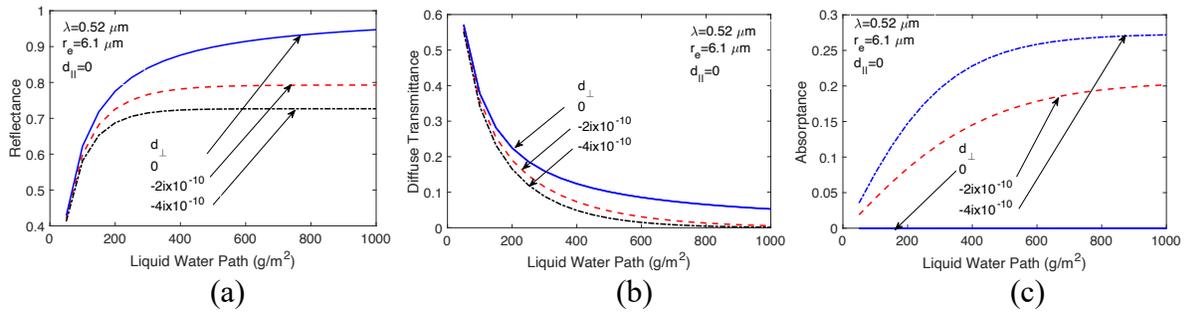

(a) (b) (c)

**Figure 4 Cloud properties: liquid water path dependence.** (a) reflectance (albedo), (b) diffuse transmittance, and (c) absorptance as a function of the liquid water path at given parameters.

Figure 5(a)-(c) show the cloud properties as a function of the effective radius. For a given liquid water path, the smaller is the effective radius, the more is the surface area, and the larger is the absorptance. The reflectance and transmittance reduce correspondingly.

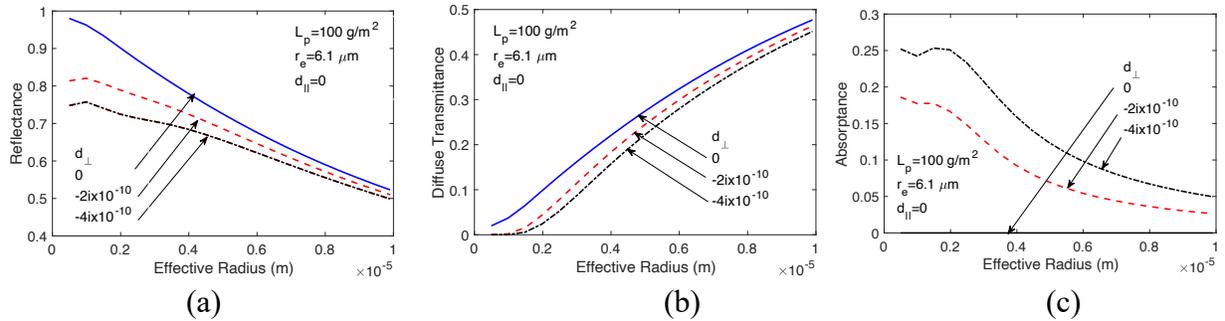

(a) (b) (c)

**Figure 5 Cloud properties: effect radius dependence.** (a) reflectance (albedo), (b) diffuse transmittance, and (c) absorptance as a function of the effective radius at given parameters.

**d. Discussion**

At this stage, a direct comparison with experiments is not practical due to uncertainties in the Feibelman parameters and their wavelength dependence. The above calculation however, motivates the following discussion on the past literature.

Cloud models have consistently predicted that the short wavelength (solar spectral) cloud forcing, defined as the net difference between the cloudy-sky (all-sky) and clear-sky net downward radiation (difference between downward and upward), should be the same on top of the clouds ($C_t$) and on the earth surface ($C_s$), i.e., $\gamma=C_s/C_t\sim 1$, because the absorption difference between water droplets and water vapor is small for the same water mass. Experimentally, however, this ratio is often found to be ~1.5 and even larger values have been reported (Cess et al. 1995; Ramanathan et al. 1995; Pilewskie and Valero 1995). One method used to arrive at this ratio is by plotting the cloud albedo as a function of the transmittance as Cess et al. (1995) did using satellite data and Pilewskie and Valero (1995) did using flights data. Figure 6 shows the total short-wave albedo vs. the total transmittance, calculated from the spectral reflectance and transmittance data averaged by AM1.5 solar insolation. It can be seen from the figure that the values of the slope changes from s=-0.91 when Im($d_\perp$)=0 to s=-0.63 when Im($d_\perp$)=-4x10$^{-10}$. Cess et al. (1995) found an experimental value of s=-0.59, while the climate models gave s=-0.8. Thus, including photomolecular effect makes s value smaller and can potentially led to agreement with experiments, should more accurate values of the Feibelman parameters be determined.

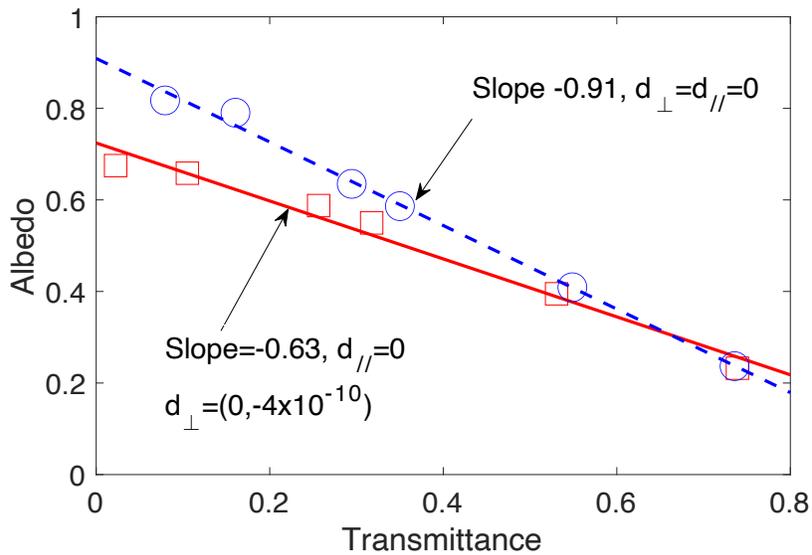

**Figure 6 Albedo vs. transmittance.** Both albedo and transmittance are calculated from the spectral data weighted by AM1.5 spectrum. The inclusion of surface absorption leads to a smaller slope value. Slope values inferred from experiments range from -0.48 to -0.61.(Cess et al. 1995; Pilewskie and Valero 1995).

## 4. Summary

In summary, the author suggests here that the recently discovered photmolecular effect might provide an important piece of physics that has been missing in the cloud models. Including this new absorption mechanism can potentially explain the anomalous cloud absorption: many measured or inferred cloud absorptance values are higher than current climate models could predict.

The author develops an approach to simulate the potential photomolecular effect on cloud absorption. The treatment is built on the modified Mie solutions using the generalized boundary conditions for the Maxwell equations based on the Feibelman parameters. This approach enables us to calculate the cloud absorption and extinction coefficients including the photomolecular effect, which were subsequently feed into the ERT for cloud property calculations. Solution of the ERT shows that the surface absorption arising from the photomolecular effect is important in the visible and near infrared wavelength range. Depending on the values of the Feibelman parameters, the effective radius of the droplets, and the liquid water path, the photomolecular effect can lead to a large absorption in the clouds. Since clous absorption is the most uncertain part of climate models, this work suggests that further assessment of the phomolecular effect on cloud absorption anomaly is warranted.

**Acknowledgments:**

The author would like to thank Ms. Caterina Grossi for bringing my attention to the cloud absorption anomaly and her initial effort in simulation the photomolecular effect using Monte Carlo method. The author is grateful for the encouragement from Dr. William Collins. This work is funded by a MIT Bose Award. Support from MIT J-WAFS and UMRP exploring the desalination applications of the photomolecular effect are also helpful in the author's current pursue to understand the photomolecular effect.

**Availability Statement:**

Inhouse software developed by the author to solve ERT is available upon reasonable requests.

**References**

Baldi, A., 2024: Mie_Scattering_and_Absorption_Sphere. *https://github.com/andrea-baldi/Mie_Scattering_and_Absorption_Sphere/releases/tag/v1.0.2,.*

Cess, R. D., and Coauthors, 1995: Absorption of solar radiation by clouds: observations versus models. *Science (1979)*, **267**, 496–499.

Chandrasekhar, S., 1960: *Radiative transfer*. Dover.

Chen, G., 2024: Modeling photomolecular effect using generalized boundary conditions for Maxwell equations. *Commun Phys*, **7**, 330, https://doi.org/10.1038/s42005-024-01826-z.

Craig F. Bohren, C. F., and D. R. Donald R. Huffman, 1998: *Absorption and Scattering of Light by Small Particles*.

Danielson, R. E., D. R. Moore, and H. C. van de Hulst, 1969: The transfer of visible radiation through clouds. *Journal of the Atmosphere Sciences*, **26**, 1078–1087.

Feibelman, P. J., 1982: Surface electromagnetic fields. *Prog Surf Sci*, **12**, 287–408.

Fritz, S., and T. H. Macdonald, 1951: Measurements of absorption of solar radiation by clouds. *Bulletin American Meteorological Society*, **32**, 205–209.

Gonçalves, P. A. D., T. Christensen, N. Rivera, A. P. Jauho, N. A. Mortensen, and M. Soljačić, 2020: Plasmon–emitter interactions at the nanoscale. *Nat Commun*, **11**, 366, https://doi.org/10.1038/s41467-019-13820-z.

Hale, G. M., and M. R. Querry, 1973: Optical constants of water in the 200-nm to 200-μm wavelength region. *Appl Opt*, **12**, 555–563, https://doi.org/10.1364/AO.12.000555.

Hewson, E. W., 1943: The reflection, absorption, and transmission of solar radiation by fog and cloud. *Quarterly Journal of the Royal Meteorological Society*, **69**, 47–62, https://doi.org/10.1002/qj.49706929808.

Kindel, B. C., P. Pilewskie, K. S. Schmidt, O. Coddington, and M. D. King, 2011: Solar spectral absorption by marine stratus clouds: Measurements and modeling. *Journal of Geophysical Research Atmospheres*, **116**, D10203, https://doi.org/10.1029/2010JD015071.

Lai, Y. C., S. Q. Chen, L. Y. Mou, and Z. N. Wang, 2021: Nanoscale electromagnetic boundary conditions based on Maxwell's equations. *Acta Physica Sinica*, **70**, 230301, https://doi.org/10.7498/aps.70.20211025.

Landry, M. J., C. Fu, J. H. Zhang, J. Li, G. Chen, and M. Li, 2025: Theory of the Photomolecular Effect. *ArXiv*, 2501.08373, https://doi.org/https://doi.org/10.48550/arXiv.2501.08373.

Lee, Y. G., J. Kim, Y. Jung, H. K. Cho, J. Kim, and J. H. Koo, 2021: Cloud Impacts on Korea Shortwave Radiation Budget: Estimation from a Deterministic Model with Surface Measurements. *Asia Pac J Atmos Sci*, **57**, 321–330, https://doi.org/10.1007/s13143-020-00196-0.

Li, Z., H. W. Barker, and L. Moreaut, 1995: The variable effect of clouds on atmospheric absorption of solar radiation. *Nature*, **376**, 486–490.

——, T. P. Ackerman, W. Wiscombe, and G. L. Stephens, 2003: Have clouds darkened since 1995? *Science (1979)*, **302**, 1151–1152, https://doi.org/10.1126/science.302.5648.1151.

Li, Z., W. Wiscombe, G. Stephens, and T. Ackerman, 2004: Disagreements over cloud absorption - response. *Science (1979)*, **305**, 1240–1240, https://doi.org/10.1126/science.305.5688.1240.

Liebsch, A., 1997: *Electronic excitations at metal surfaces*.

Liou, K., 1973: A numerical experiment on Chandrasekhar's discrete-ordinate method for radiative transfer: applications to cloudy and hazy atmospheres. *Journal of Atmospheric sciences*, **130**, 1303–1326.

Lv, G., Y. Tu, J. H. Zhang, and G. Chen, 2024: Photomolecular effect: visible light interaction with air-water interface. *Proc Natl Acad Sci*, **121**, e2320844121.

Modest, M. F., 2013: *Radiative heat transfer*. 3rd ed. Academic Press,.

O'Hirok, W., and C. Gautier, 2003: Absorption of shortwave radiation in a cloudy atmosphere: Observed and theoretical estimates during the second Atmospheric Radiation Measurement Enhanced Shortwave Experiment (ARESE). *Journal of Geophysical Research: Atmospheres*, **108**, https://doi.org/10.1029/2002jd002818.

Oreopoulos, L., A. Marshak, and R. F. Cahalan, 2003: Consistency of ARESE II cloud absorption estimates and sampling issues. *Journal of Geophysical Research: Atmospheres*, **108**, AAC13, https://doi.org/10.1029/2002jd002243.

Pilewskie, P., and F. P. J. Valero, 1995: Direct observations of excess solar absorption by clouds. *Science (1979)*, **267**, 1626–1629.

Ramanathan, V., B. Subasilar, G. J. Zhang, W. Conant, R. D. Cess, J. T. Kiehl, H. Grassi, and L. Shi, 1995: Warm pool heat budget and shortwave cloud forcing: a missing physics? *Science (1979)*, **267**, 499–503.

Schaich, W. L., and W. Chen, 1989: Nonlocal corrections to Fresnel optics: How to extend d-parameter theory beyond jellium models. *Phys Rev B*, **39**, 10714–10724, https://doi.org/10.1103/PhysRevB.39.10714.

Schmidt, K. S., and Coauthors, 2010: Apparent absorption of solar spectral irradiance in heterogeneous ice clouds. *Journal of Geophysical Research Atmospheres*, **115**, https://doi.org/10.1029/2009JD013124.

Sommer, A. P., K. F. Hodeck, D. Zhu, A. Kothe, K. M. Lange, H. J. Fecht, and E. F. Aziz, 2011: Breathing volume into interfacial water with laser light. *Journal of Physical Chemistry Letters*, **2**, 562–565, https://doi.org/10.1021/jz2001503.

——, P. Schemmer, A. E. Pavláth, H.-D. Försterling, Á. R. Mester, and M. A. Trelles, 2020: Quantum biology in low level light therapy: death of a dogma. *Ann Transl Med*, **8**, 440–440, https://doi.org/10.21037/atm.2020.03.159.

Stamnes, K., 1986: The theory of multiple scattering of radiation in plane parallel atmospheres. *Reviews of Geophysics*, **24**, 299–310, https://doi.org/10.1029/RG024i002p00299.

Stamnes, K., S.-C. Tsay, W. Wiscombe, and K. Jayaweera, 1988: Numerically stable algorithm for discrete-ordinate-method tradiative transfer in multiple scattering and emitting layered media. *Appl Opt*, **27**, 2502–2509.

Stamnes, K., W. Wiscombe, and I. Laszlo, 2000: *DISORT, a general-purpose Fortran program for discrete-ordinate-method radiative transfer in scattering and emitting layered media: documentation of methodolody*.

Stephens, G. L., 1978: Radiation profiles in extended water clouds. I: theory. *J Atmos Sci*, **35**, 2111–2121.

——, 1996: How much solar radiation do clouds absorb? *Science (1979)*, **271**, 1131–1133.

Stephens, G. L., and S. -C Tsay, 1990: On the cloud absorption anomaly. *Quarterly Journal of the Royal Meteorological Society*, **116**, 671–704, https://doi.org/10.1002/qj.49711649308.

Trenberth, K. E., J. T. Fasullo, and J. Kiehl, 2009: Earth's global energy budget. *American Meteorological Society*, 311–324, https://doi.org/10.1175/2008BAMS2634.I.

Tu, Y., J. Zhou, S. Lin, M. Alshrah, X. Zhao, and G. Chen, 2023: Plausible photomolecular effect leading to water evaporation exceeding the thermal limit. *Proc Natl Acad Sci U S A*, **120**, e2312751120, https://doi.org/10.1073/pnas.2312751120.

Valero, F. P. J., and Coauthors, 2003: Absorption of solar radiation by the clear and cloudy atmosphere during the Atmospheric Radiation Measurement Enchanced Shortwave Experiments (ARESE) I and II: Observations and models. *Journal of Geophysical Research D: Atmospheres*, **108**, AAC9, https://doi.org/10.1029/2001jd001384.

Valero, F. P. J., R. D. Cess, and S. K. Pope, 2004: Disagreements Over Cloud Absorption. *Science (1979)*, **305**, 1239–1240, https://doi.org/10.1126/science.305.5688.1239.

Verma, G., V. Kumar, A. Kumar, and W. Li, 2024: Unveiling photon-driven nonlinear evaporation via liquid drop interferometry. *Opt Lett*, **49**, 4074–4077, https://doi.org/10.1364/ol.527346.

Wild, M., A. Ohmura, H. Gllgen, and E. Roeckner, 1995: Validation of general circulation model radiative fluxes using surface observations. *J Clim*, **8**, 1309–1321.

Zender, C. S., B. Bush, S. K. Pope, A. Bucholtz, W. D. Collins, J. T. Kiehl, F. P. J. Valero, and J. Vitko, 1997: Atmospheric absorption during the Atmospheric Radiation Measurement (ARM) Enhanced Shortwave Experiment (ARESE). *J Geophys Res*, **102**, 29901–29915.

Zhang, J. H., R. Mittapally, G. Lv, and G. Chen, 2025: Superthermal solar interfacial evaporation is not due to reduced latent heat of water. *Energy Environ Sci*, https://doi.org/10.1039/D4EE05591H.

Zhang, M., R. D. Cess, and X. Jing, 1997: Concerning the interpretation of enhanced cloud shortwave absorption using monthly-mean Earth Radiation Budget Experiment/Global Energy Balance Archive measurements. *Journal of Geophysical Research Atmospheres*, **102**, 25899–25905, https://doi.org/10.1029/97jd02196.

Zhao, F., and Coauthors, 2018: Hierarchically Nanostructured Gels. *Nat Nanotechnol*, **13**, 489–496, https://doi.org/10.1038/s41565-018-0097-z.

Zhen, B., S. G. Johnson, C. W. Hsu, J. Lee, J. D. Joannopoulos, M. Soljačić, and S.-L. Chua, 2013: Observation of trapped light within the radiation continuum. *Nature*, **499**, 188–191, https://doi.org/10.1038/nature12289.